\documentclass[journal]{IEEEtran}
\usepackage{cite}
\ifCLASSINFOpdf
\else
\fi
\usepackage{amsmath}
\usepackage{graphicx}
\usepackage{amsfonts}
\usepackage{xcolor}
\usepackage{algorithm}
\usepackage{algpseudocode}
\usepackage{array} % required for text wrapping in tables
\usepackage{authblk}

\usepackage{breqn}
\begin{document}
\bstctlcite{IEEEexample:BSTcontrol}
\title{DRL-based Power Allocation in LiDAL-Assisted RLNC-NOMA OWC Systems}

{\author[1*]{Ahmed A. Hassan\thanks{\textsuperscript{*}Email: ahmed.4.hassan@kcl.ac.uk}}}
\author[1]{Ahmad Adnan Qidan}
\author[1]{Taisir Elgorashi}
\author[1]{Jaafar Elmirghani}

\affil[1]{Department of Engineering, King's College London, London,  United Kingdom.}
        
% <-this % stops a space
%\thanks{M. Shell was with the Department
%of Electrical and Computer Engineering, Georgia Institute of Technology, Atlanta,
%\thanks{}


\markboth{ Submitted to XXX Jan~2026}%
{Shell \MakeLowercase{\textit{et al.}}: Bare Demo of IEEEtran.cls for IEEE Journals}



\maketitle

\begin{abstract}Non-orthogonal multiple access (NOMA) is a promising technique for optical wireless communication (OWC), enabling multiple users to share the optical spectrum simultaneously through the power domain. However, imperfect channel state information (CSI) and residual decoding errors deteriorate NOMA performance, especially in realistic dense-user indoor scenarios. In this work, we model an OWC system that integrates light detection and localization (LiDAL) and random linear network coding (RLNC) within a NOMA framework. LiDAL exploits spatio-temporal information to improve user CSI, while RLNC enhances data resilience in the successive decoding process, resulting in a LiDAL-assisted RLNC-NOMA OWC system. Power allocation (PA) is crucial in this system due to complex interactions between multiple users and the coding and detection processes, but optimizing continuous PA dynamically can be computationally prohibitive. To address this, we adopt a deep reinforcement learning (DRL) framework to efficiently learn near-optimal PA strategies. In particular, a DRL-based normalized advantage function (NAF) algorithm is proposed to maximize the average sum rate, and its performance is compared to deep deterministic policy gradient (DDPG), gain ratio PA (GRPA), and exhaustive search. The results indicate that NAF closely matches exhaustive search, is 39\% faster than DDPG, and improves the average sum rate by 4.6\% over GRPA, while accounting for user location estimation errors.
\end{abstract}


% \fi
\IEEEpeerreviewmaketitle

\section*{Introduction}
\IEEEPARstart{O}{}ptical wireless communication (OWC) has the potential to deliver high-speed connectivity that enables the growing data demand of users for data-intensive applications such as video streaming, immersive reality, and artificial intelligence (AI).
OWC systems can make use of existing indoor lighting based on commercially available  light-emitting diodes (LEDs) or laser diodes (LDs) as transmitters, together with photodetectors (PDs) as receivers, allowing multigigabit data rates to be achieved \cite{pathak_visible_2015,11164782}. To make this technology more accessible and efficient,  non-orthogonal multiple access (NOMA) was proposed as a  multiple access technique (MA) that enables several users to share the same resource simultaneously across different domains such as power, frequency, time, code, and space \cite{feng_multiple_2019}. Compared with conventional orthogonal MA schemes such as  OFDMA, TDMA, and CDMA, NOMA offers several advantages, including enhanced user fairness, improved spectral efficiency, and higher achievable data rates \cite{maraqa_survey_2020,qidan2025}. 

As a direct consequence of these features, power-domain NOMA, well established within OWC, requires highly accurate channel state information (CSI) to perform critical operations in NOMA-based transmission, including user grouping or pairing, power allocation, and signal superposition. However, assuming static channel conditions over a given time period, or modeling them using standard probabilistic distributions such as Rayleigh or Gaussian for CSI estimation, is unrealistic in practical indoor OWC scenarios \cite{wang_location-based_2020, 9064520,8636954}. The uncertainty in CSI estimation arises from various factors, including user mobility, shadowing, and quantization errors.
 Based on this, it becomes evident that quantifying the effect of imperfect CSI on the performance of NOMA-based OWC systems is essential \cite{tran_performance_2022}. In \cite{al-hameed_lidal_2019,hassan_lidal_assisted_2025}, a light detection and localization system (LiDAL) was introduced to exploit location-based information as an auxiliary source and improve the accuracy of CSI estimation. This system passively captures the spatio-temporal presence and location of users, allowing channel estimates to be updated more accurately and frequently.
{The LiDAL system considered in this work is fundamentally different from RF-based localization systems, as the nonlinearity of optical sources and the limited eye-safety transmit power jointly reshape localization operation and its achievable accuracy. In contrast, in RF-based localization systems, multipath, co-channel interference, and signal-strength fluctuations are the primary challenges \cite{9705498}}.

 To characterize CSI imperfections, the Cramer–Rao lower bound (CRLB) was also employed in these studies as an unbiased error estimator. Nevertheless, even with improved CSI, residual errors in NOMA interference cancellation can still cause potential degradation in the useful data throughput of NOMA systems.
In \cite{hassan_lidal_assisted_2025, hassan_random_2023-1}, 
random linear network coding (RLNC) was integrated with NOMA in an OWC system. In the combined RLNC-NOMA scheme, the optical access point encodes data packets as random linear combinations of the original packets over superimposed NOMA signals. At the receiver side, as long as a sufficient number of independent combinations are received, the original information can be reliably reconstructed, thereby mitigating the propagation of errors in the information carried by superimposed NOMA user signals.

In general, NOMA asymmetrically allocates optical power among users, giving more power to those with poor channel conditions and less power to those with better channels. It also allows users to correctly decode their intended signals using techniques such as successive interference cancellation (SIC). Therefore, obtaining accurate CSI for each user and determining the optimal set of power ratios or coefficients is essential to balance performance metrics such as reliability, throughput, and fairness across all users in the OWC system. Several power allocation (PA) techniques were proposed in NOMA-based OWC systems and mainly focus on maximizing the capacity of the OWC system in terms of the ergodic sum rate and the number of users served. Such strategies include fixed PA (FPA) \cite{kizilirmak_non-orthogonal_2015}, gain ratio PA (GRPA) \cite{marshoud_non-orthogonal_2016}, exhaustive search PA \cite{yin_performance_2015}, normalized gain difference PA \cite{chen_performance_2018}, and max-min fairness with sum rate maximization PA \cite{shen_optimal_2017}. 
Moreover, different methods are also proposed to investigate the power allocation strategy, user fairness, and energy efficiency. However, most of the reported contributions to such joint resource allocation problems in NOMA-based OWC systems are NP-hard \cite{yang_fair_2017}\cite{maraqa_achievable_2021}, especially when the user CSI is highly fluctuated and/or correlated in scenarios involving mobility or dense indoor networks. Therefore, obtaining an optimal solution is challenging due to the high level of uncertainty and the significant computational complexity involved. \cite{9839259,10279546}. Heuristic and meta-heuristic optimization techniques, such as genetic algorithms (GA) \cite{wang_subcarrier_2021} and simulated annealing (SA) \cite{maraqa_achievable_2021}, have been proposed to suboptimally solve NP-hard optimization problems, which may converge to local optima. Consequently, the performance of NOMA can be limited in various scenarios, particularly in multi-parameter, realistic, real-time indoor environments. Therefore, it is crucial to develop efficient and low-complexity practical techniques to achieve optimal solutions in such realistic scenarios.

Machine learning (ML), and specifically reinforcement learning (RL), a subset of ML, enables a computing node, i.e., the central unit in a wireless communication system, to mimic human cognitive processes and intelligent behaviors to solve complex problems. RL is a framework in which an intelligent agent learns to interact with an environment, making a series of sequential decisions to effectively achieve specific goals. Q-learning (off-policy) and SARSA (on-policy) are well-known RL algorithms that have attracted significant attention from both industry and academia for solving challenging optimization problems in various networking systems \cite{sutton_reinforcement_2018}. Both algorithms rely on a Q-table to store the optimal sequence of actions that maximize future rewards. Moreover, they have demonstrated substantial benefits in applications such as channel prediction \cite{gaballa_study_2023}, user scheduling \cite{ding_resource_2021}, power allocation, and energy efficiency optimization \cite{zhang_dynamic_2018}. In the context of OWC, several studies have applied these RL algorithms to enhance system performance in various aspects, as reported in \cite{da_silva_noma-based_2020,elgamal_q-learning_2021,long_q-learning_2023}. However, Q-learning and SARSA are not suitable for large-scale optimization problems, such as those in MISO or MIMO network architectures, because they must explore and learn the entire continuous state-action space. This requirement leads to enormous Q-table storage and intensive lookup operations, making it challenging to converge to the optimal policy.

To overcome these limitations, deep learning, specifically deep neural networks, has been integrated into RL, forming deep reinforcement learning (DRL) frameworks such as policy gradient (PG), actor-critic (AC), and deep Q-network (DQN). DQN combines Q-learning with deep neural networks to handle complex environments with large state spaces and high-dimensional discrete actions. Additionally, advanced DRL algorithms such as deep deterministic policy gradient (DDPG) \cite{lillicrap_continuous_2019} and normalized advantage functions (NAF) \cite{gu_continuous_2016} are designed for problems with continuous action spaces and large-scale, multidimensional state spaces, where both are model-free and off-policy. However, DDPG converges slowly compared to NAF and may not always reach the Nash equilibrium, as reported in \cite{gu_continuous_2016, sun_distributed_2023}.

In contrast to the literature discussed above, this work leverages LiDAL-assisted location information and RLNC encoding. The proposed framework improves CSI estimation accuracy and enhances NOMA performance. To efficiently optimize system performance in complex scenarios, a NAF-based DRL model is introduced, capable of handling continuous action spaces and high-dimensional state spaces, enabling more effective power allocation compared to other conventional and RL-based methods.

The main contributions of this work are summarized as follows:

\begin{itemize}
    \item A LiDAL-assisted RLNC-NOMA OWC system is modeled to exploit location-based information for enhanced CSI estimation and encodes data packets as random linear combinations, improving NOMA performance in OWC.
\item A complex optimization problem is formulated with the aim of maximizing the average sum rate of the system and enhancing user experience. The optimization problem is then reformulated, and a NAF-based DRL model is introduced to efficiently solve it in dynamic scenarios, overcoming the limitations of traditional RL algorithms in large state-action spaces.
\item The accuracy of user location estimation and its impact on network performance are analyzed, and the performance of the proposed NAF algorithm is evaluated against DDPG, GRPA, and exhaustive search power allocation schemes in terms of sum rate, average user data rate, and user fairness.
\end{itemize}

The remainder of this article is organized as follows. First, we provide the configuration of the setup of the proposed DRL-based NAF agent, analyze its learning complexity, and evaluate its performance within the proposed system model compared to benchmark power allocation schemes. Next, we present our discussion and remarks. Finally, we detail the system model of the proposed RLNC-NOMA downlink transmission assisted by the MIMO-based LiDAL system. In this context, we formulate the power allocation problem for the proposed system model and describe the DRL-based NAF algorithm and agent architecture that is used to solve it.

\begin{figure*}[h!tbp]
    \centering
    \includegraphics[width=0.8\linewidth]{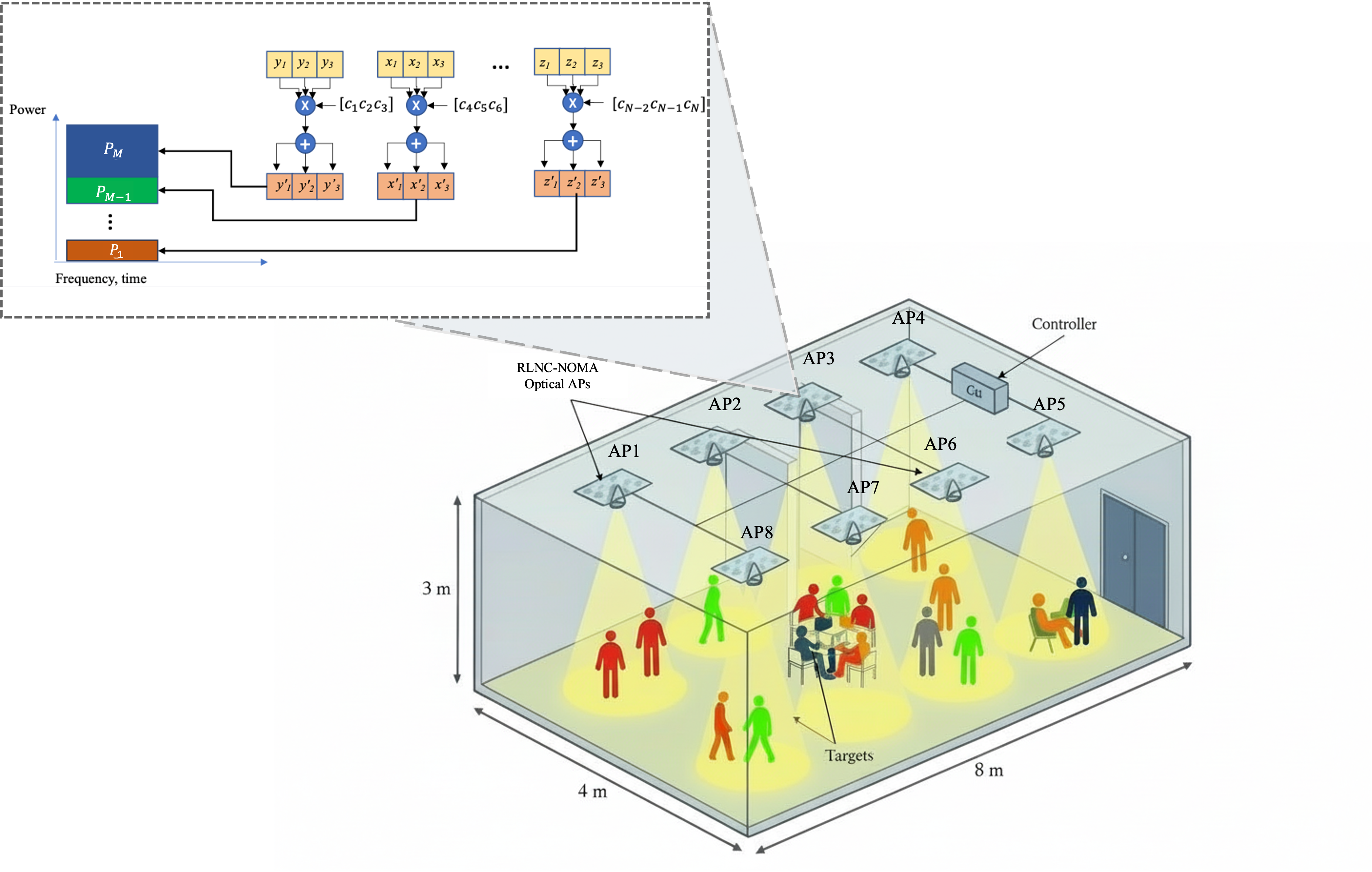}
    \caption{LiDAL-based RLNC-NOMA system model.}
    \label{fig:LiDAL-RLNC-NOMA-system-model}
\end{figure*}

\section*{Results}
In this section, we introduce the DRL-based environment setup and present a detailed performance analysis of the proposed DRL-based NAF agent. This agent interacts with the designed environment to maximize the average sum rate by optimizing the power allocation among the grouped communication users. The optimization process employs the suboptimal LiDAL-based grouping scheme within the MDP framework formulated to solve the non-convex power allocation problem of the proposed LiDAL-assisted RLNC-NOMA system. Furthermore, we provide a comprehensive comparison of the results obtained using our approach with those from the DDPG agent, as well as conventional methods such as GRPA and grid-based exhaustive search based power allocation techniques.

\subsection*{System Environment Setup}
In this setup, we consider the OWC environment of the proposed system model to be an indoor study space. This space is equipped with eight optical APs (i.e., $K=8$), which together form eight overlapping optical footprints, as illustrated in Fig.~\ref{fig:LiDAL-RLNC-NOMA-system-model}. The maximum number of users that the MIMO-LiDAL system can detect and localize, based on the environment settings and parameters specified in Table~\ref{tab:NAF-parameters}, is derived from equation (88) in \cite{al-hameed_lidal_2019}. Consequently, the maximum user capacity within the defined indoor OWC environment is set to $N=32$ users.  

Without loss of generality, we assume that these users are uniformly distributed and randomly located (restricted by the minimum resolution $\Delta R = 0.3$ m and the placement of furniture) within each LiDAL-based NOMA group associated with optical AP $k$. { We adopt a snapshot‑based model aligned with the nomadic behavior of the users in both LiDAL localization and NOMA downlink time slots and change discretely between frames, capturing slow indoor mobility as a sequence of stationary snapshots.}

 Each group $k$ accommodates 4 users (i.e., $M_k=4, \;\forall k \in K$). Moreover, each AP $k$ is assigned a dedicated optical bandwidth to avoid channel correlation between overlapping user groups. Each user is equipped with a single PD receiver with a narrow FOV angle (i.e., $\Psi_c = 40^\circ$), and the minimum RLNC-NOMA data rate requirement per user is $\gamma_{\min}=100$ Mbps \cite{dong_adaptive_2019}. Furthermore, the initial power allocation factor for users in group $k$, denoted as $\alpha_i^k$ for $i=\{1,2,3,4\}$ and $k=\{1,2,\dots,8\}$, is set equally to 0.25 at the initial state $s_{t=0}$ \cite{hammadi_deep_2022}. To interact with this environment, DRL-based NAF and DDPG agents are implemented and trained on a computing node of the university HPC cluster, equipped with 32~GB of RAM and two 16-core 3.0~GHz processors. Both agents are implemented in the Python programming language using the TensorFlow library.

\subsection*{DRL-based NAF Agent Setup}
{The NAF agent consists of target and evaluation networks that share the same ANN architecture but serve different roles. Both networks are optimized and updated using the ADAM optimization algorithm, which provides adaptive learning for non-stationary objectives and noisy data. The architecture of both networks includes an input layer corresponding to the environment state space, which comprises the estimated channel gains, location noise coefficients, and the allocated power factors at any given time. This input layer is followed by three hidden layers with 16 neurons each, representing the state-value function $V(s)$. The output layer of the $V(s)$ network produces a scalar value for each state $s$. The mean-action function $\mu(a)$ (policy network) shares the same input layer as the $V(s)$ network but is connected to three separate hidden layers, which produce outputs corresponding to the power allocation coefficients as a vector of actions.
Finally, the quadratic scaling advantage function network $A(s,a)$ combines the observed environment states with the output of the mean-action $\mu(a)$ policy function (i.e., the optimal actions). This network has three hidden layers with 32 neurons each, reflecting its increased complexity. Its output layer represents the lower-triangular matrix used to construct the positive semi-definite matrix $\mathbf{P}(s)$ in the advantage function (\ref{eq:advantage-fun}). All hidden layers employ the rectified linear unit (ReLU) activation function for non-linear transformations, while linear activation functions are used in the output layers to enable continuous action scaling. The hyperparameters of the proposed DRL-based NAF agent are summarized in Table~\ref{tab:NAF-parameters}.} {These values are obtained by manual heuristic tuning around the standard NAF defaults from \cite{hammadi_deep_2022}. Specifically, the batch size \(I\), soft update factor \(\tau_Q\), and replay buffer size \(D\) are adjusted to ensure minimal episode reward variance while exceeding the total throughput achieved by the conventional GRPA method on validation episodes across multiple random seeds.}

\begin{table}[h!]
    \centering
        \caption{DRL-based NAF agent hyperparameters.}

    \begin{tabular}{|>{\raggedright\arraybackslash}p{0.5\linewidth}|>{\centering\arraybackslash}p{0.3\linewidth}|}
        \hline
        \textbf{Parameter} & \textbf{Value}  \\
         \hline
        Discount Factor, \(\varsigma\) & 0.995 \cite{hammadi_deep_2022}  \\
        \hline
        Learning rate, & 0.0001 \cite{hammadi_deep_2022}  \\
        \hline
        Target NN soft update factor, \(\tau_Q\) & 0.001  \\
        \hline
        OU(\( \theta,\mu,\sigma\)) & (0.15, 0, 0.3)  \\
        \hline
        Replay buffer size, \(D\) & 10000 \\
        \hline
        mini batch size, \(I\) & 32 \\
        \hline
        \(V(s)\),\(\mu(s)\) and \(P(s)\) NNs width  & 16, 16, 32  \\
        \hline
         Number of output actions & 4  \\
        \hline
         Activation function & ReLU  \\
        \hline
        Number of hidden layers& 3  \\
          \hline
          Gradient clipping & 1  \\
           \hline
    \end{tabular}
    \label{tab:NAF-parameters}
\end{table}
\subsection*{DRL-based NAF Learning Complexity}
The proposed DRL-based NAF algorithm follows the steps outlined in Algorithm~\ref{alg:DRL-NAF-training} to learn the optimal power allocation policy for RLNC-NOMA users in the LiDAL-based group $k$, which is connected to an optical AP $k \in K$, in order to achieve the highest average sum rate of the overall OWC system. In this process, the NAF agent’s evaluation network interacts with the indoor OWC environment to collect experience tuples $e_t = (s_t, a_t, r_t, s_{t+1})$ and store them in a replay buffer $\mathcal{D}$. At each training iteration, a mini-batch of transitions of size $I$ is randomly sampled from this buffer, and target Q-values are computed using the agent's target network. The parameters of the target network are updated by minimizing the mean squared error between the target and predicted Q-values via gradient descent.  

To analyze the overall training complexity, the evaluation network has an action dimension of $4K$. Therefore, the cost of each training step can be expressed as
\begin{align}
    O(I(L_s Z^2 + 16K^2)),
\end{align}
where $L_s$ denotes the number of hidden layers per network, and $Z$ denotes the number of neurons per layer.  

For comparison, the cost of training the same action dimension using the DDPG algorithm is
\begin{align}
    O(I(2 L_s Z^2 + 4 K)),
\end{align}
noting that two updates are required per training step for the actor and critic networks in DDPG. Despite the quadratic dependence on the action dimension in the NAF algorithm, it is more efficient and achieves faster convergence than DDPG due to the use of two identical network architectures with shared weights, which simplifies the training process.  

For a fair comparison, it is assumed that both NAF and DDPG networks have the same number of hidden layers and neurons per layer in the performance analysis.

{On the other hand, the computational cost of using a grid-based exhaustive search to converge to the optimal power allocation for all 4-users in the \(K\) groups can be expressed as
\begin{align}
    O\big(K \; I_{s} \;M_{\text{grid}}^4\big),
\end{align}
where \(I_s\) denotes the random combinatorial samples per group and \(M_{\text{grid}} = ((\alpha_{\max} - \alpha_{\min})/\Delta_s) + 1\) denotes the number of grid points of the power allocation variable per user. This formulation directly reflects the discretization of the continuous power allocation interval \([\alpha_{\min}, \alpha_{\max}]\) with a fixed step size \(\Delta_s\). As result, the search space grows exponentially with the number of grid points per dimension, leading to a complexity that is highly sensitive to both the resolution of the grid (i.e., the value of \(\Delta_s\)) and the number of users per group. Therefore, even though grid-based exhaustive search guarantees finding the globally optimal solution within the fine-grain discrete power domain, its complexity increases exponentially with the number of users in each group, making it impractical for realistic NOMA-based systems in scenarios with high user density.}

\subsection*{Implementation Performance Benchmark}
{We further evaluate the implementation efficiency of the proposed DRL-based NAF algorithm and compare it against the DDPG algorithm and an exhaustive search baseline in terms of training time, memory demands, and convergence time. All methods are evaluated under an identical computing environment and with the same training hyperparameters to ensure the fairness of the comparison between the NAF- and DDPG-based implementations.
The results in Table \ref{tab:performance-benchmark} indicate that the proposed DRL-based NAF algorithm is the most efficient approach overall, achieving the lowest training time per iteration (0.27 s) and the fastest convergence time (2.98 s). Compared to DDPG, it reduces the iteration time by approximately 39\% and the convergence time by approximately 56\% in our system model, indicating a clear implementation advantage. However, the NAF agent architecture requires a slightly higher memory demands than DDPG, due to the additional parameters required by its quadratic advantage function structure.}

\subsection*{DRL-based NAF Learning Evaluation}
\subsubsection{Learning Convergence}
\begin{figure}
    \centering
    \includegraphics[width=1.0\linewidth]{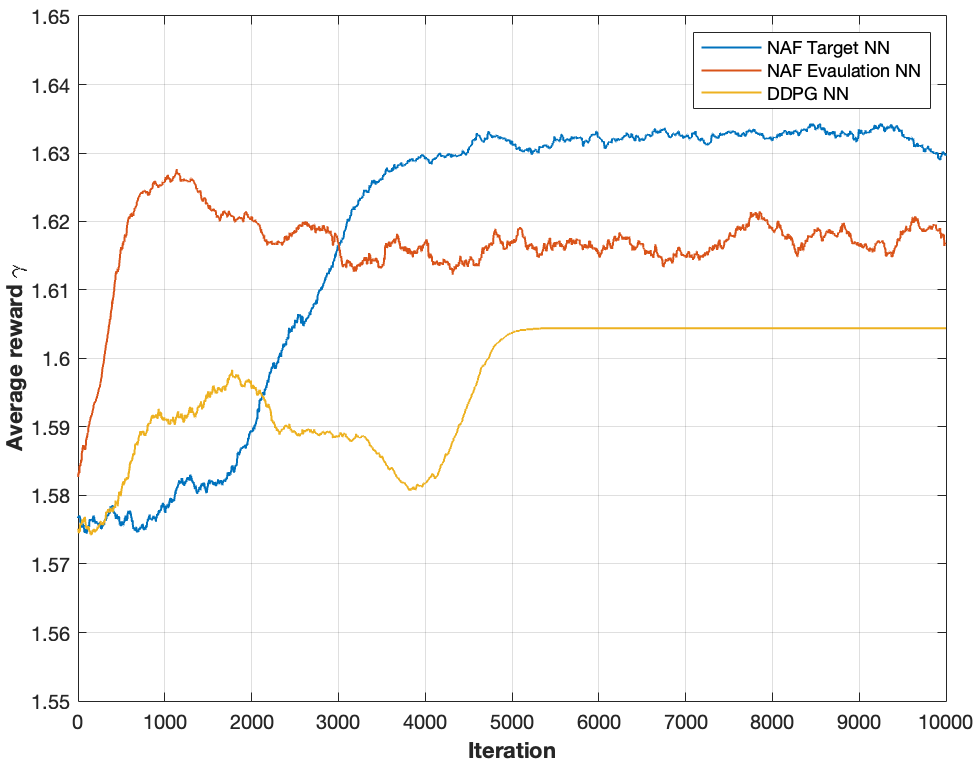}
    \caption{DRL-based NAF vs DDPG learning convergence. }
    \label{fig:DRL-NAF-convergance}
\end{figure}

Fig. \ref{fig:DRL-NAF-convergance} illustrates the interaction between the proposed DRL-based NAF agent and the LiDAL-based OWC environment, compared to the DDPG agent during the learning phase. {
During each iteration, all user locations associated to a LiDAL-based NOMA group \(k\) are considered fixed while the agents (i.e., NAF, DDPG) interact with the environment. Across iterations, the user locations are re-sampled from the set of feasible locations within the same LiDAL-based group \(k\), so the training experience independent realizations of user locations and corresponding CSI, reflecting a sequence of independent stationary snapshots that capture user nomadic mobility behavior.} The graph reflects the speed of learning and policy convergence (i.e., selection of the optimal power allocation factors) over iterations. It also shows the progressive updates of the NAF agent's target network relative to its evaluation network, and the stabilization of both networks’ convergence over time (around 3000 iterations), where the NAF agent outperforms the DDPG agent. Specifically, the evaluation and target networks achieve average sum rates of 1.617 Gbps and 1.631 Gbps, respectively, corresponding to the average reward. In contrast, the DDPG network achieves a lower average sum rate of 1.597 Gbps, taking into account the negative rewards (penalties) incurred during the agent’s interaction with the environment.
\subsubsection{Learning Loss}
\begin{figure}[h!tbp]
    \centering
    \includegraphics[width=1.0\linewidth]{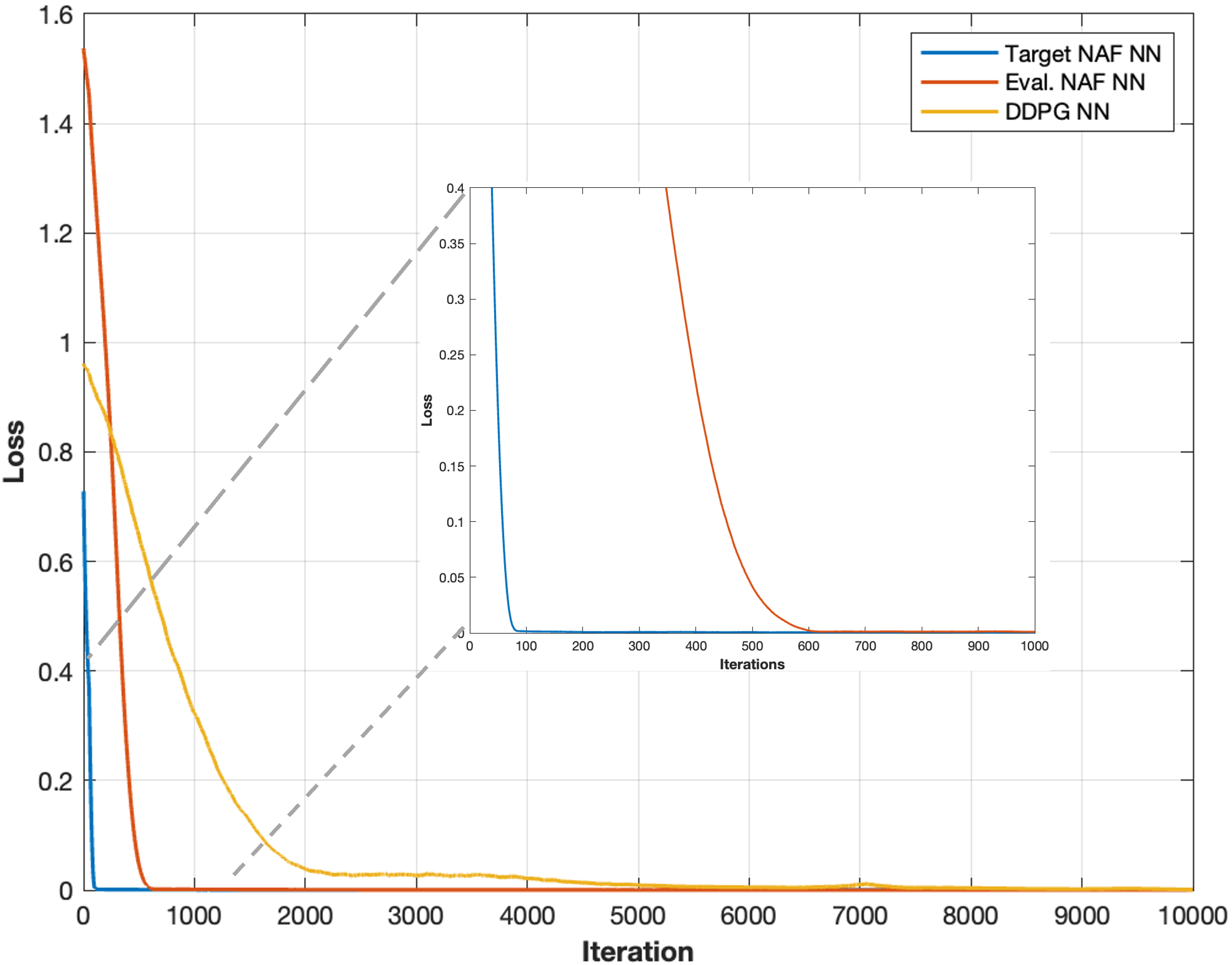}
    \caption{DRL-based NAF vs DDPG loss function evaluation.}
    \label{fig:DRL-NAF-loss}
\end{figure}

As illustrated in Fig. \ref{fig:DRL-NAF-loss}, the loss function, representing the difference between the predicted and target Q-values, is evaluated for both the NAF and DDPG algorithms during the learning process. The results demonstrate the efficiency of the proposed DRL-based NAF algorithm compared to DDPG, as it more effectively minimizes the error and rapidly improves the accuracy of the estimated optimal power allocation factors in the proposed LiDAL-based OWC system.

\begin{table}[h!]
    \centering
        \caption{Implementation performance benchmark }

    \begin{tabular}{>{\centering\arraybackslash}p{0.1\linewidth}>{\centering\arraybackslash}p{0.3\linewidth}>{\centering\arraybackslash}p{0.2\linewidth}>{\centering\arraybackslash}p{0.2\linewidth}}
    \hline
         Method & Training/search time per iteration (sec) & Average Memory usage (MB) & Convergence time (sec) \\
    \hline
         NAF&  0.27 & 359.7 & 2.98\\
    \hline
         DDPG & 0.44 & 352.14 & 6.76\\
    \hline
    Exhaustive search & 9.23 & 425.70& 8.14\\
    \hline
    \end{tabular}
    \label{tab:performance-benchmark}
\end{table}
\subsection*{System Groups Average Sum Rate}

Here, we discuss the optimization of power allocation within the proposed LiDAL-based OWC system by employing DRL-based NAF and DDPG agents, and we compare their performance to conventional GRPA and exhaustive search-based PA methods. The GRPA scheme represents a conventional NOMA approach, in which power allocation factors are determined based on the users with the best channel conditions \cite{tao_performance_2018}. Specifically, in our case, the user with the highest location order ($i=1$) is assigned to an optical AP $k$, where $k \in \{1,2,\dots,8\}$.  

\begin{figure}[h!tbp]
    \centering
    \includegraphics[width=1.0\linewidth]{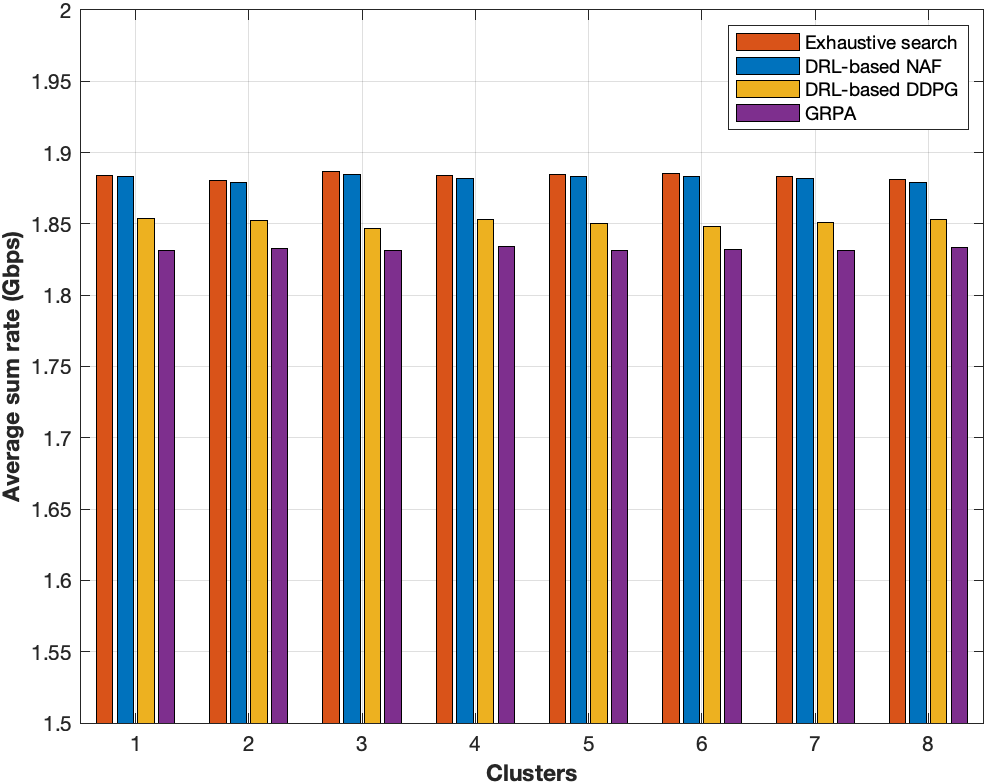}
    \caption{The average sum rate of LiDAL-NOMA OWC user groups.}
    \label{fig:LiDAL-NOMA-sumrate-avg-cls}
\end{figure}

The results in Fig.~\ref{fig:LiDAL-NOMA-sumrate-avg-cls} indicate that the proposed DRL-based NAF power allocation technique achieves performance comparable to that of the exhaustive search PA method. Additionally, both NAF and DDPG agents outperform the conventional GRPA scheme in terms of average sum rate, which was previously applied as a baseline in the analysis of the proposed LiDAL-based NOMA user grouping scheme \cite{hassan_lidal_assisted_2025}. Specifically, the application of the NAF agent improves the average sum rate of the proposed LiDAL-assisted RLNC-NOMA system by 1.64\% compared to the DDPG agent and 4.6\% compared to the GRPA scheme, resulting in an average sum rate of 1.882~Gbps.

\subsection*{System Average User Data Rate}
\begin{figure}[h!tbp]
    \centering
    \includegraphics[width=1.0\linewidth]{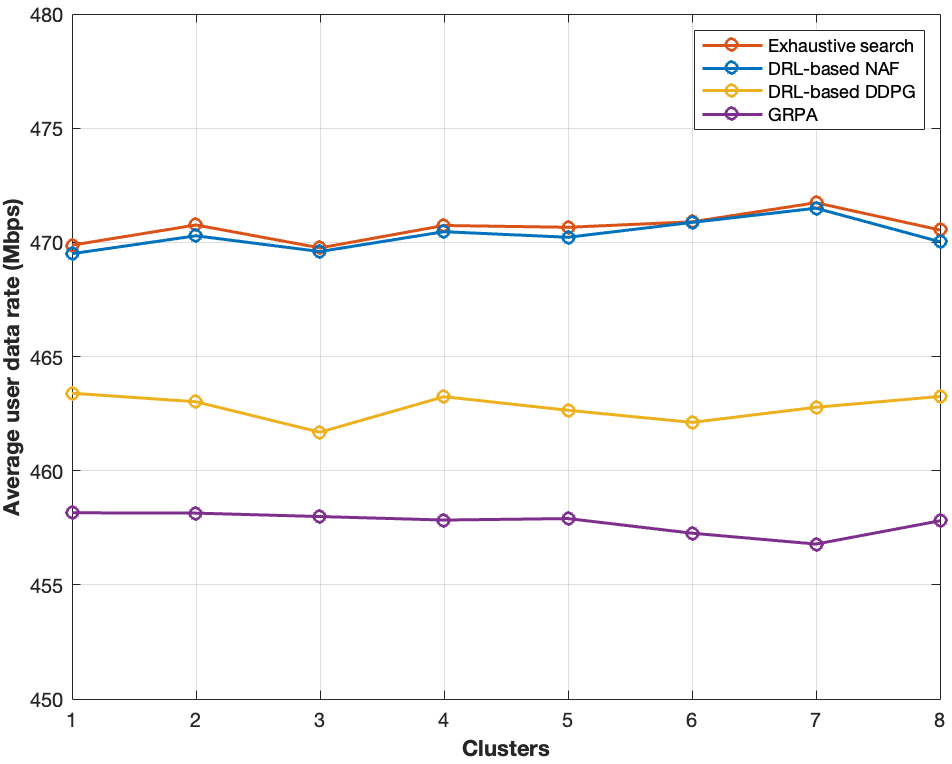}
    \caption{The average LiDAL-NOMA user data rate per group.}
    \label{fig:LiDAL-NOMA-cls-user-rate}
\end{figure}

From the user perspective, we evaluated the average achievable data rate within each LiDAL-based user group, and the results are depicted in Fig.~\ref{fig:LiDAL-NOMA-cls-user-rate}. The highest average user data rate of 471.7~Mbps is observed in group~7, where users experience optimal channel conditions and minimal localization errors due to reduced interference and fewer signal reflections from walls and other opaque objects relative to their positions in the environment. In contrast, users in group~3 record the lowest average data rate of 469.5~Mbps, which can be attributed to their locations being surrounded by a larger number of corners and wall surfaces. This increases signal reflections, resulting in higher localization errors (i.e., lower localization probability) and, consequently, reduced accuracy of channel state estimates for these users.
 
\subsection*{User Date Rates Fairness}
\begin{figure}[h!tbp]
    \centering
    \includegraphics[width=1.0\linewidth]{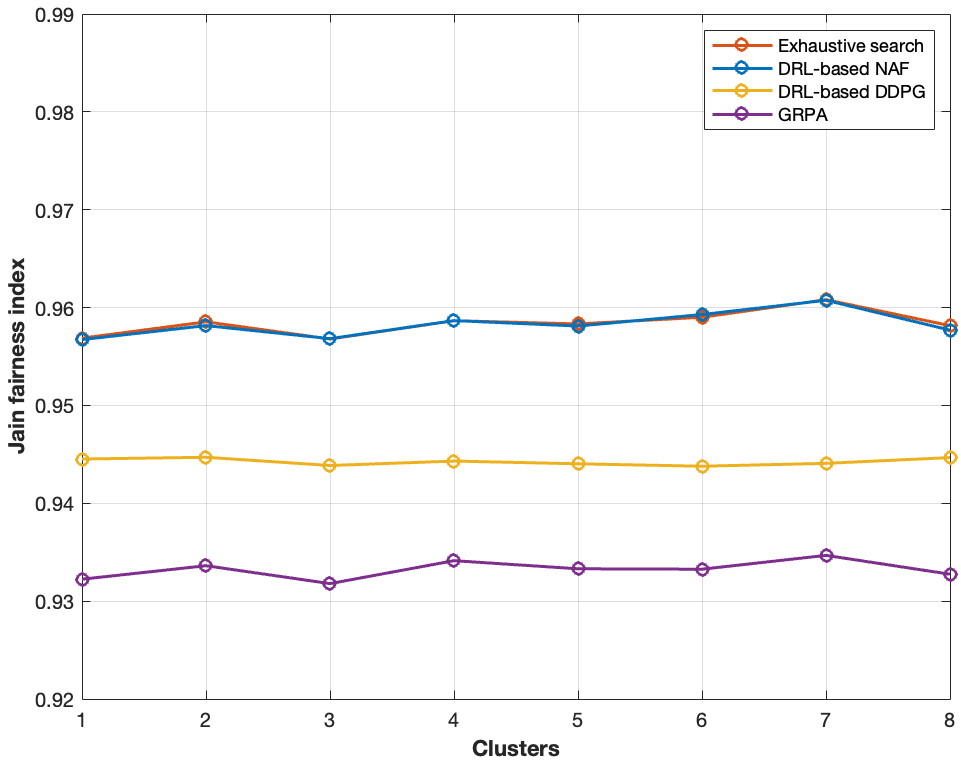}
    \caption{Jain's fairness index of the user data rates per group.}
    \label{fig:LiDAL-NOMA-cls-user-fairness}
\end{figure}

{We applied Jain's fairness index as a standard metric to measure how uniformly the data rates are distributed among users within each LiDAL-based NOMA group. The results presented in Fig. \ref{fig:LiDAL-NOMA-cls-user-fairness} demonstrate that both the DRL-based NAF algorithm and the exhaustive search achieve the highest level of fairness in the data rates among users across all eight groups formed in the LiDAL-based NOMA system. Although the DDPG algorithm achieves a slightly lower level of fairness compared to NAF and exhaustive search methods, it maintains a consistent fairness value of 94.5\% across all groups as it dynamically adjusts power for varying user locations and nomadic behavior, ensuring minimum data rate constraints. Furthermore, the results clearly confirm the impact of localization errors and the quality of CSI estimates on the fairness index of users within groups 3 and 7. This fairness analysis confirms that the proposed DRL-based NAF algorithm not only improves system performance in terms of sum rate but also ensures a highly uniform distribution of data rates among users.}

\subsection*{Power Allocation vs. Location Error}

In this part, we extend our analysis to examine the relationship between the average power allocation ratio and the location estimation error within the proposed LiDAL-assisted RLNC-NOMA system. This analysis employs the proposed DRL-based NAF power allocation technique and compares the results with those obtained using the three aforementioned schemes. The results indicate that both the DRL-based NAF and exhaustive search PA methods tend to allocate higher power ratios to users located in positions with the highest average location estimation errors. It is important to note that the user location error is not the only factor affecting the estimated CSI channel coefficients. Other factors, such as the user access distance (i.e., the LOS distance between the user's receiver and the associated optical AP), the Euclidean distance between the locations assigned to a particular optical AP, and interference from other RLNC-NOMA users, are also considered. 

Furthermore, the results show that the optimized power allocation using the DRL-based NAF closely matches that of the exhaustive search PA method, taking into account the estimated user location errors. It should also be noted that noise in the proposed realistic environment, along with reflections from walls and opaque objects, negatively impacts the location estimation accuracy in the LiDAL-assisted RLNC-NOMA system.

\begin{figure}[h!tbp]
    \centering
    \includegraphics[width=1.0\linewidth]{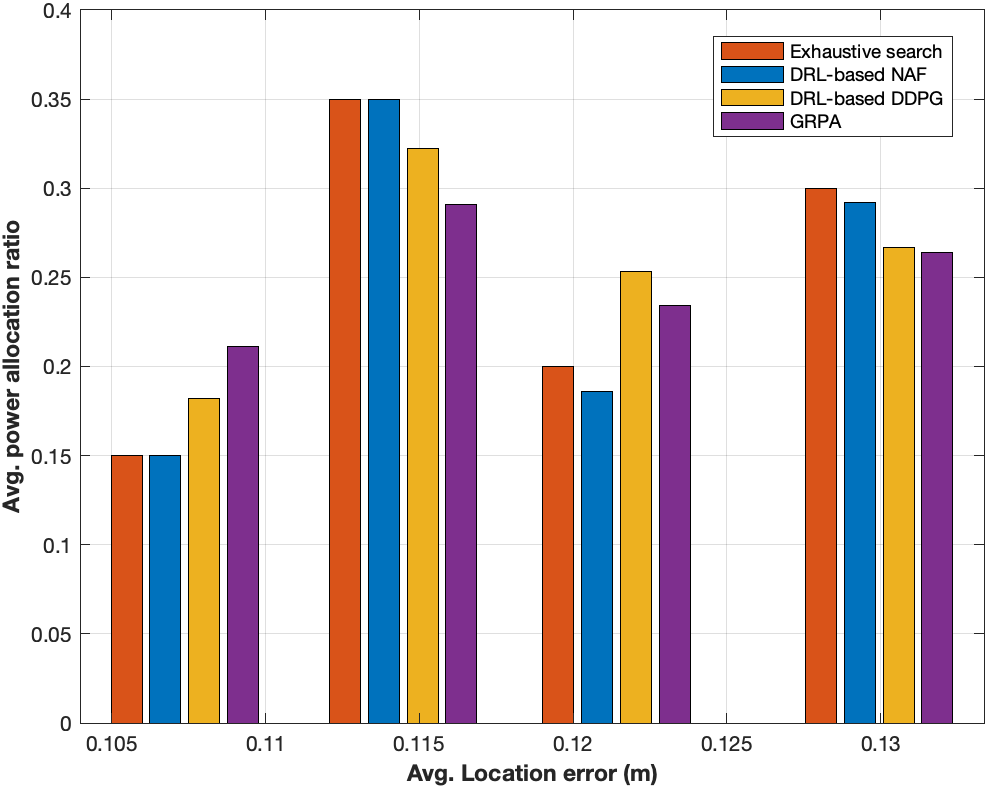}
    \caption{Average power allocation vs location estimation error per user.}
    \label{fig:LiDAL-NOMA-error-PA}
\end{figure}

\section*{Discussion}
The article proposes a LiDAL-assisted RLNC-NOMA OWC system that uses location information and random linear network coding to improve CSI estimation and NOMA performance in realistic indoor environments. It formulates a complex optimization problem to maximize average sum rate and user fairness and introduces a NAF-based DRL model to efficiently solve it in dynamic, high-dimensional, continuous-action settings. The results show that imperfect CSI and location errors significantly challenge dense-user indoor OWC, but the NAF algorithm converges faster than DDPG and yields higher average sum rates than conventional NOMA schemes. It closely approximates optimal power allocation under varying CSI and location uncertainties, achieving performance near exhaustive search and demonstrating robustness, efficiency, and practical suitability for indoor OWC systems.

\section*{Methods}
{In this section, we present the system model of the LiDAL-based RLNC-NOMA framework, formulate the power allocation problem within this system context, and describe in detail the DRL-based NAF algorithm and its agent architecture, that are proposed to solve the reformulated power allocation problem.}

\subsection*{System Model}
We consider a LiDAL-based RLNC-NOMA system model as illustrated in Fig.1. This system consists of two integrated systems: MIMO-LiDAL system for user detection and localization and RLNC-NOMA system for robust user access and communication. Both are connected to a central control unit (i.e., controller).

The MIMO-LiDAL system utilizes a set of \(K\) optical APs to transmit short optical pulses within dedicated time slots, while ceiling-mounted LiDAL receivers periodically capture reflections from a set of \(N\) users with nomadic behavior (i.e., alternating between mobility and stationary), as well as from other opaque objects in the indoor environment (i.e., walls, tables, doors). {We assume user locations are randomly selected from a uniform spatial distribution over the accessible area (subject to furniture and localization resolution constraints), and users remain stationary for the duration of each LiDAL probing time slot and corresponding NOMA downlink frame. Across frames, user locations change independently according to this distribution, modeling slow indoor mobility with piecewise‑stationary channel realizations.}

Each LiDAL receiver detects the presence of users by applying cross-correlation processing and estimates the time of arrival (TOA) of the detected user signals. The system operates in both monostatic (transmitter and receiver located in the same optical AP) and bistatic (transmitter and receiver located in separate optical APs) configurations, enabling the acquisition of multiple independent range measurements for each detected user within a predefined time slot. A central controller processes these ranges to obtain user location estimates and the associated localization error for each user. These output parameters are later used by the controller to define the user location ordering, AP association for data transmission and to derive the corresponding CSI estimation errors.

The RLNC-NOMA system uses the same set of optical APs to transmit superimposed NOMA signals. Each optical AP \(k\) serves a number of users with simultaneous NOMA downlink access. Before signal superposition phase, the optical AP applies RLNC to a generation of fixed-size original packets intended for an associated user. These packets are multiplied by random coefficients over the Galois field GF(\(2^8\)) and linearly combined to form coded packets. When the optical receiver of that user obtains its intended signal via SIC, it extracts the corresponding coded packets. Once a sufficient number of linearly independent coded packets have been received, the user applies matrix-based decoding operations to recover the original generation of packets without sequence synchronization and/or packet loss feedback.

For the integrated system model described above, Each optical AP \(k\) consists of a uniform array of homogenized RGB-LDs using an in-front diffuser that emits a Lambertian radiation pattern that is utilized for both communication and localization. In optical downlink access, we consider the LOS component of channel gain (i.e., ideal CSI) between an arbitrary user \(i\) and the associated optical AP \(k \in K\) to be modeled by
\begin{multline}
      h_i = \\
    \begin{cases}
        \dfrac{(m+1)A_{PD}}{2\pi {\Lambda_i^k}^2}cos^m(\phi_i)cos(\psi_i)T_f(\psi_i)g_c(\psi_i),\; 0\leq \psi_i \leq \Psi_c\\
        0, \quad \! \text{otherwise}
    \end{cases} \\
\end{multline}
where \(\phi_i\) and \(\psi_i\) denote the irradiance and incidence angles between the optical AP \(k\) and the PD receiver of the user \(i\), respectively. \(\Lambda_i^k\) denotes the measured access distance between the location of the user \(i\) and its associated optical AP \(k\) which corresponds in particular to the LOS distance between the PD receiver of the user \(i\) and the optical AP \(k\). \(\Psi_c\) represents the FOV angle of the PD receiver and \(A_{PD}\) denotes its physical area, \(m = - \frac{ln2}{ln\,cos{{\Phi}_{1/2}}}\) is the Lambertian index of optical AP \(k\) corresponding half-power semi-angle denoted by \(\Phi_{1/2}\). Note that, \(T_f(\psi_i)\) and \(g_c(\psi_i)\) denote the optical filter gain of the PD receiver and its optical concentrator gain, respectively.

The integrated MIMO-LiDAL system is not only capable of locating users, but can also derive additional information from the user location within the realistic indoor environment considered.  
Therefore, we assume that the location \(X_i\) of a user \(i\) can be characterized by the CRLB of location error \(B^L_i\) and the measured LOS distance \( \Lambda_{i}^k\) between the PD receiver of the user \(i\) and its associated optical AP \(k\). Therefore,
the corresponding imperfect CSI estimated error \(( \Delta h_i^*= h_i - h_i^*)\) of the user \(i\) at that location \(X_i\) can be estimated by \cite{ma_robust_2023}
\begin{align}
    \Delta h_i^* \simeq \Delta_c \bigg(\frac{\eta^{m+1}}{({\Lambda_i^k}^2 + {B^L_i}^2)^{\frac{m+3}{2}}} - \frac{\eta^{m+1}}{({\Lambda_i^k}^2)^{\frac{m+3}{2}}}\bigg),
    \label{eq:LiDAL-NOMA-CSI-approx}
\end{align}
where \(\Delta_c = \frac{(m+1)A_{PD}}{2\pi}T_f(\psi)g_c(\psi)\), \(\eta\) denotes the height difference between a user \(i\) and the ceiling of the room that is assumed to be constant due to the limited height variation of all users in the indoor environment considered \cite{al-hameed_lidal_2019}.

 We assume that there are \(K\) user groups (and hence \(K\) optical APs), indexed by \(k \in \{1, \dots, K\}\). Each group \(k\) consists of \(M_k\) users associated with AP \(k\). These user groups are formed using a grouping method that aligns with NOMA principles \cite{hassan_lidal_assisted_2025} by exploiting localization information obtained from the proposed MIMO-LiDAL system. The optical signal received by an arbitrary user \(i, i \in M_k\), assigned to optical AP \(k\) can be expressed by \cite{hassan_lidal_assisted_2025}
\begin{align}
   \mathcal{Y}_{i} = \mathfrak{R}_i (h_i^* + \Delta h_i^*) \mathcal{P}_t\sum_{n=1}^f \alpha_i  c_{in}.b_{in} + n_i ,\;
\end{align}

where \(\mathfrak{R}_i\) denotes the responsivity of the PD receiver of user \(i\). \(h_i^*\) denotes the estimated channel gain coefficient, \(\Delta h_i^*\) is the corresponding error in the estimate of the channel coefficient derived in (\ref{eq:LiDAL-NOMA-CSI-approx}). \(\mathcal{P}_t\) represents the total optical power that can be transmitted by the corresponding optical AP \(k\). The NOMA power allocated to the user \(i\) is indicated by the coefficient \(\alpha_i\). 
\(c_{in}\) denotes an RLNC coefficient chosen from the finite Galois field GF\((2^8)\). \(b_{in}\) is a set of original packets \(f\) that are transmitted over the superimposed optical signal. (\(n_i = n^*_i + n^'_i\)) represents the noise generated by the user's PD receiver \(n^*_i\) and the additional noise due to imperfect SIC \(n^'_i\). This additional noise is attributed to the noisy and/or correlated CSI estimation of the user \(i\).

 In the NOMA downlink scenario, each optical AP \(k\) has an exclusive communication bandwidth \(B^k_w\) dedicated to a RLNC-NOMA downlink group \(k\).  In each group \(k\), the location order of associated users \(M_k\) is defined using the LiDAL-based NOMA grouping scheme. Thus, it is assumed that the power allocation factors \(\alpha^k_1,\ \alpha^k_2,\cdots,\alpha^k_{M_k}\) are assigned to the sorted users \(i=1,2,3,\cdots,M_k\), respectively. Suppose an arbitrary user \(i\) in location \(X_i\) and the remaining users in lower location order \(m^*\), where \(( i+1\le m^*\le M_k)\). This user \(i\) detects and removes interfered signals from respective users of lower location order (i.e., \(m^*>i\)) by applying the SIC protocol. Consequently, the achievable rate of the user \(i\) associated with optical AP \(k\) can be derived from
\cite{maraqa_achievable_2021}\cite{wang_tight_2013}.
\begin{multline}
\gamma^k_i = \\
\textstyle B^k_w log_2 \bigg( 1 + \dfrac{e}{2\pi}\dfrac{\mathfrak{R}_i^2(h^*_i+\Delta h^*_i)^2{\alpha^k_i}^2 P_t^2}{\sum^{i-1}_{j=1}\mathfrak{R}_i^2(h^*_i+\Delta h^*_i)^2{\alpha^k_j}^2 P_t^2 + \sigma^2_t}\bigg),
 \\i=2,3,\cdots,M_k
\label{eq:LiDAl-NOMA-user-i-drate}
\end{multline}
Hence, \(\dfrac{e}{2\pi}\) is used to satisfy the optical signal constraints of intensity modulation/direct detection (IM/DD) of the optical AP \(k\). \(\sigma_t\) denotes the standard deviation for a real-valued noise modeled as an additive white Gaussian with zero mean \(\mathcal{N}(0,\sigma_t)\). 

For users with a higher location order \(m^\prime\) (\(1\le m^\prime < i \)) associated with optical AP \(k\), user \(i\) treats their signals (i.e., \(m^\prime < i\)) as noise. Thus, the achievable rate of the user \(i\) to detect a user \(m^\prime\) information can be expressed by
\begin{multline}
   \textstyle \gamma^k_{m^\prime \rightarrow i} =\\ B^k_w log_2 \bigg( 1 + \dfrac{e}{2\pi}\dfrac{\mathfrak{R}_i^2(h^*_i+\Delta h^*_i)^2{\alpha^k_{m^\prime}}^2 P_t^2}{\sum^{i-1}_{j=m^\prime}\mathfrak{R}_i^2(h^*_i+\Delta h^*_i)^2{\alpha^k_j}^2 P_t^2 + \sigma^2_t}\bigg),
    \\i=2,3,\cdots,M_k
    \label{eq:signal-separation-rate}
\end{multline}
 In the case of the highest location order, the interference is canceled from all other users by the SIC process. Therefore, the user's achievable rate at the highest location order (i.e., \( i=1\)) is given by
\begin{align}
\gamma^k_1 = B^k_w log_2 \bigg( 1 + \dfrac{e}{2\pi}\dfrac{\mathfrak{R}_1^2(h^*_1+\Delta h^*_1)^2{\alpha^k_1}^2 P_t^2}{\sigma^2_t}\bigg).
\label{eq:LiDAl-NOMA-user-1-drate}
\end{align}

\subsection* {Power Allocation Problem Formulation}
\label{sec:power-allocation-problem}
In this subsection, we formulate the power allocation problem of the proposed system model for a highly dense indoor environment. The objective is to maximize the average sum rate of the users that are grouped and ordered according to NOMA principles, utilizing the location-based information provided by the MIMO-LiDAL system.
Accordingly, the optimal power allocation problem of the proposed LiDAL-based RLNC-NOMA system can be given by
\begin{align}
\max \; \dfrac{1}{K} \sum^K_{k=1}\sum^{M_k}_{i=1} \gamma^k_i, \quad \forall i \in M_k , \forall k \in K
\label{eq:objective-fun}
\end{align}
s.t.
\begin{align}
    \sum^{M_k}_{i=1}\alpha^k_i = 1,\quad \forall i \in M_k, k \in K 
    %\tag{\ref{eq:objective-fun}.1}
    \label{eq:const1}
\end{align}
\begin{align}
    0 \leq \alpha^k_i \leq 1,\quad \forall i \in M_k, k \in K 
    %\tag{\ref{eq:objective-fun}.2}
     \label{eq:const2}
\end{align}
\begin{align}
       \textstyle \alpha^k_1 \leq \alpha^k_2 \leq \cdots \leq \alpha^k_i \leq \alpha^k_{i+1} \leq ... \leq \alpha^k_{M_k},\quad \forall i \in M_k, k \in K 
       %\tag{\ref{eq:objective-fun}.3}
        \label{eq:const3}
\end{align}
\begin{align}
    \gamma^k_i \geq \gamma_{min}, \quad \forall i \in M_k, k \in K. 
    %\tag{\ref{eq:objective-fun}.4}
     \label{eq:const4}
\end{align}
where \(\gamma_{min}\) denotes the minimum achievable rate required \cite{hassan_lidal_assisted_2025} for the cancellation of RLNC-NOMA interference in (\ref{eq:signal-separation-rate}) and \(\gamma^k_i\) denotes the achievable data rate of the user \(i\) and is defined according to (\ref{eq:LiDAl-NOMA-user-i-drate}) for \(i=2,3,\cdots,M_k\) and (\ref{eq:LiDAl-NOMA-user-1-drate}) for the user of the highest location order (\(i=1\)) within a group \(k\) that is associated with optical AP \(k\).

Typically, the power allocation problem formulated above can generally be assumed stationary and solved using the conventional non-convex optimization techniques \cite{9064520,qidan_resource_2021}. However, these techniques become more complex when CSI variability is introduced in the context of realistic user mobility and nomadic behaviors in non-stationary OWC environments. Moreover, the real-time location-based CSI derived by the MIMO-LiDAL system requires a real-time optimization strategy to effectively handle dynamic channel conditions. Therefore, the RL-based optimization approach is considered an effective solution to this problem within the context of our proposed user-dense RLNC-NOMA system model.

\subsection* {RL-based Framework}
\begin{figure}[h!tbp]
    \centering
    \includegraphics[width=0.7\linewidth]{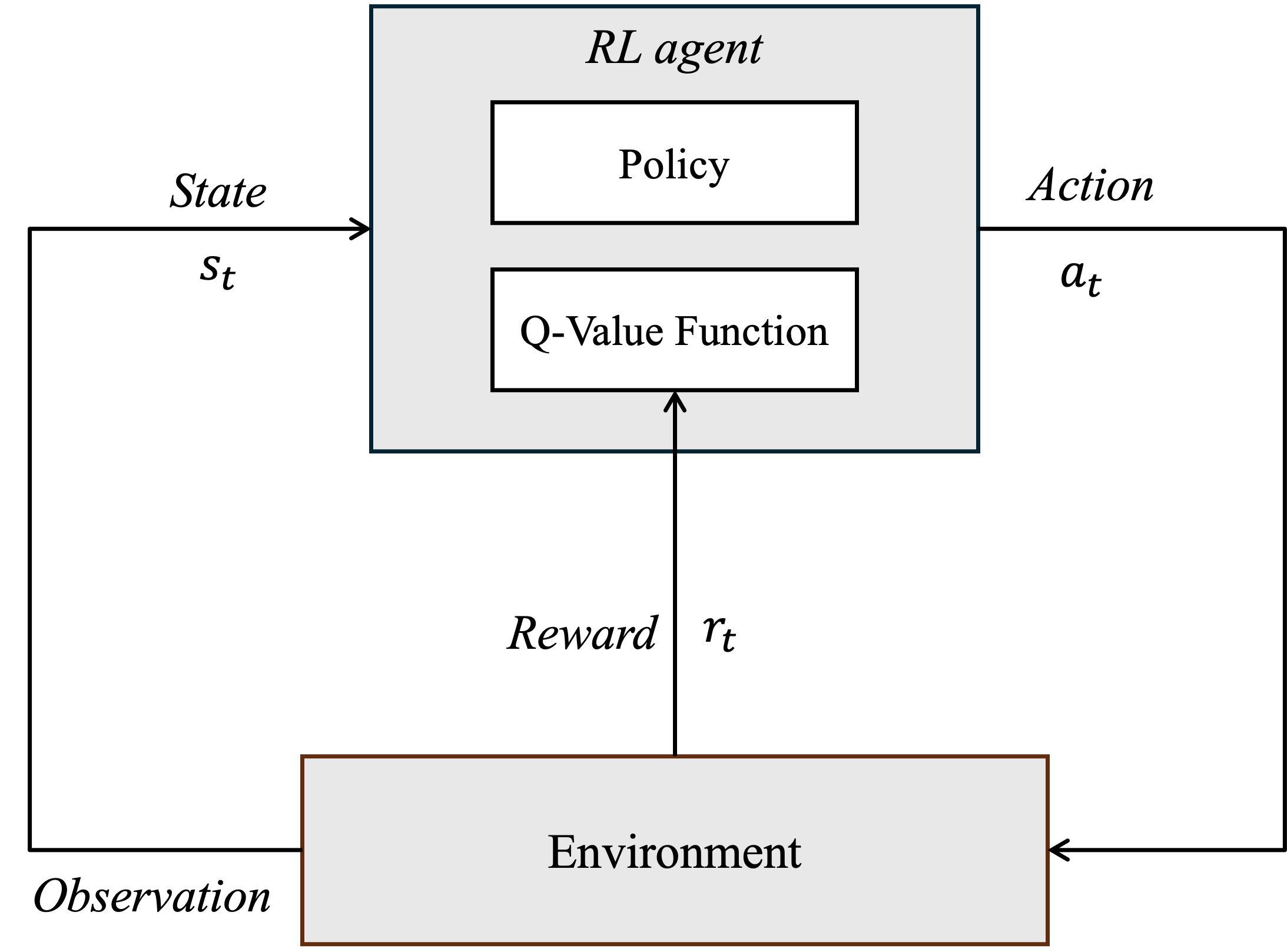}
    \caption{Schematic architecture of RL framework.}
    \label{fig:RL-architecture}
\end{figure}

 Generally, in the RL framework, an agent interacts with the environment and selects an action for this environment, and the status changes after the environment accepts that action. At the same time, a feedback is propagated to the agent (reward or penalty), as illustrated in Fig.\ref{fig:RL-architecture}. Ideally, the agent selects the next action based on the expected future reward and the current state of the environment. The RL agent tries to find a strategy or policy that maximizes the accumulative rewards in a series of best actions over time. At a glance, the key elements of the RL framework are the following. 

\begin{itemize}
    \item \textbf{States:} the state \(s_t \in S\) represents the information observed about the environment at any given time \(t\) (i.e., the location of the user, the channel gain between the user and the associated AP,\dots,etc.).
    \item \textbf{Actions:} the action is what the agent interacts with the environment and is denoted as \(a_t\). In continuous space, actions are vectors of real values (i.e., user power allocation factors).
    \item \textbf{State-action values:} denoted by Q-function \(Q_\pi(s_t,a_t)\) which estimates the expected return (cumulative reward) of taking a specific action \(a_t\) in a specific state \(s_t\) and following a certain policy \(\pi\).
    \item \textbf{Policy:} denoted by \(\pi(.)\) is the mapping between the actions to be taken by the agent in any given state \(s_t\) of the environment. 
    \item \textbf{Rewards:} denoted by \(r_t\), which represents the outcome (i.e., the average achievable sum rate) obtained immediately after an agent takes a specific action \(a_t\) in a given state \(s_t\) at time \(t\) that leads to state \(s'\).
\end{itemize}
At time \(t'\), this system generates an experience tuple \(e' = (s_t,a_t,r_t,s')\), which is stored in the replay buffer \(D\). The main goal of the RL agent is to maximize cumulative future discounted rewards, following a certain optimal policy \(\pi^*\) that maps the best actions to the states of the environment by iterative approximation of the Q-value function \(Q_{\pi}(s_t,a_t)\) known as the Q-learning process. The Q-value function can be mathematically expressed by 
\begin{align}
    Q_{\pi}(s_t,a_t) = \mathbb{E}[r_t|s_t,a_t] = \mathbb{E} \bigg[\sum^{\infty}_{i=0} \varsigma^i r_{t}|s_t,a_t \bigg]
    \label{eq:q-value-discounted-rewards}
    \\
    = \mathbb{E}[r_t + \varsigma Q_{\pi}(s',a')|s_t,a_t]
\end{align}
where \(\mathbb{E[.]}\) denotes the expected value and \(\varsigma \in (0,1)\) denotes the discount factor. 
The optimal policy \(\pi^*\) is the policy that maximizes the expected Q-value function and the future cumulative rewards, which is defined as the Bellman equation and expressed by
\begin{align}
    Q_{\pi^*}(s_t,a_t) = \mathbb{E} \bigg[r_t + \varsigma   \max\limits_{a'}Q_{\pi^*}(s',a')|s_t,a_t \bigg]
    \label{eq:bellman-equation}
\end{align}
Note that, this expression divides the Q-value function into two terms: the immediate reward \(r_t\) and the future cumulative discounted rewards as denoted in (\ref{eq:q-value-discounted-rewards}), This approach decomposes the Q-value function into simpler recursive sub-problems and determine their optimal solutions. Nevertheless, the Q-value function is non-linear, so there is no analytical close-form solution to it. Consequently, iterative methods (i.e. Q-learning) have been proposed that can converge to the optimal Q-value solutions. However, these methods became impractical and computationally intensive especially in multi-user systems in continuous action- and state-space environments.

Various neural network-based approximation methods, such as DQN, DDPG and NAF have been widely adopted in the literature to estimate the Q-value function to achieve an optimal policy. DQN is a value-based method that directly approximates the Q-value function \(Q_{\pi}(s,a)\) with a neural network that structurally can only enumerate a discrete action space, which makes it unsuitable for our problem space. DDPG is primarily designed to solve optimization problems in a continuous state-action space. However, it relies on an actor-critic framework that separates policy and action-value estimations, which can lead to instability and slower convergence, especially in our precision-critical problem domain. NAF, on the other hand, is a continuous Q-learning approach with a parameterized advantage function that allows direct optimization of actions, which can lead to accurate and stable policy optimization, as will be discussed in detail in the next subsection.

\subsection*{Normalised Advantage Functions (NAF)} 

In this part, we reformulate the power allocation problem introduced earlier as a Markov decision process (MDP) to be solved using the proposed RL framework described in the previous subsection. Let \(\mathcal{A} \subseteq \mathbb{R}^{ 1 \times M}\) denote a continuous action space and \(\mathfrak{S} \subseteq \mathbb{R}^{1 \times M} \) denote a continuous state space. At each time step \(t\), a state \(s\in \mathfrak{S}\) represents the achievable data rate calculated \(\gamma^k_i\) of a user \(i, i\in M_k\), which belongs to a formed group \(k\) that is sorted and associated with optical AP \(k\). Each user \(\gamma^k_i\) is determined by the approximated CSI coefficients (\(\ h^{'}_i = h^*_i + \Delta h^*_i\)), the corresponding power allocation coefficients \(\alpha^k_i\) and the AWGN components \(\sigma^2_i \), for \(\ i=1,2,\cdots,M_k\). Consequently, the state vector \(s_t\) can be defined as 
\begin{align}
\textstyle
    s_t = (h^{'}_{1t},h^{'}_{2t},\cdots,h^{'}_{{M_k}t},\alpha^k_{1t},\alpha^k_{2t},\cdots,\alpha^k_{{M_k}t},\sigma^2_{1t},\sigma^2_{2t},\cdots,\sigma^2_{{M_k}t})
\end{align}
Note that, the estimated location and its associated CRLB error of a user \(i\) are both implied in the approximation of CSI channel coefficient \(h^{'}\) in (\ref{eq:LiDAL-NOMA-CSI-approx}). { This tight coupling between the localization error and the effective CSI enforces the RL agent to learn a policy that is robust against non‑Gaussian, geometry‑induced channel distortions rather than the purely small-scale fading as in many RF-based NOMA DRL formulations.}

for a given state \(s_t\), an agent selects an action \(a_t\) in the time step \(t\) by increasing or decreasing the power coefficient \(\alpha^k_i\) assigned to each user \(i\) associated with the optical AP \(k\). Therefore, an action \(a_t\) can be defined as 
\begin{align}
    a_t = \{(+1,-1),\cdots,(+1,-1)\}
\end{align}
where \(+1\) represents an increase in the allocated power or \(-1\) denotes a decrease in the allocated power by an optical AP \(k \in K\) for each user in the associated group of users \(M_k\). 

At each time step \(t\), an action \(a_t \in \mathcal{A}\) is taken, and the average sum rate of all users considered in \(K\) groups is calculated and denoted as an immediate reward \(r_t\) that returned for a given state \(s_t\). {Consequently, the reward function \(r_t\) can be defined as
}
{
\begin{align}
    r_t =\begin{cases}
       \Delta^{sum}_{\gamma}, & \text{if}\; 
       0 \leq \alpha^k_i \leq 1 \\ &\alpha^k_{1} \leq \cdots \leq \alpha^k_i \leq \cdots \leq \alpha^k_{M_k}\\
        -1, & \text{otherwise}
    \end{cases}
\end{align}
where 
\begin{align*}
\Delta^{sum}_{\gamma} = \sum^K_{k=1}\sum^{M_k}_{i=1} [\gamma^k_i(t) - max(0,\gamma_{min} - \gamma^k_i(t))] \\
    \forall i \in M_k , \forall k \in K
\end{align*}
}
{
Note that, the achievable data rate \(\gamma^k_i(t)\) obtained for each user \(i\) should meet the minimum achievable data \(\gamma_{min}\) (constraint \ref{eq:const4}) imposed by the RLNC-NOMA scheme. Otherwise, a penalty term will be applied. For the other two constraints (\ref{eq:const2}) and (\ref{eq:const3}), a hard negative reward will be returned.}

{In the context of the RL agent, the NAF algorithm is proposed to optimize the learned Q-function \(Q_{\pi}(s,a)\) that explicitly takes into account the joint impact of location error, optical path loss, and IM/DD constraints on the achievable rate in a continuous Q-learning process.} This optimization aims to maximize \( \max\limits_{a'}Q_{\pi}(s',a')\) in our power allocation problem domain. The Q-value function can be easily approximated during the Q-learning update by decomposing it into three components, as expressed by \cite{gu_continuous_2016}.
\begin{align}
    Q_{\pi}(s,a) = V(s) + A(s,a)
\end{align}
where \(V(s)\) is the state-value function which estimates the expected return (i.e., achievable sum rate) for a given state \(s\).
\(A(s,a)\) represents the advantage function that estimates how much better or worse an action \(a \in \mathcal{A}\) and compares it to the best action for a given state \(s\). This action is given by a vector of power allocation coefficients that are assigned to all users of the RLNC-NOMA system.

Then, the advantage function \(A(s,a)\) is parameterized as a quadratic advantage function in the action space and structured as follows \cite{gu_continuous_2016} 
\begin{align}
    A(s,a) = -\dfrac{1}{2}(a-\mu(s))^T\mathbf{P}(s)(a-\mu(s))
    \label{eq:advantage-fun}
\end{align}
where \(\mu(s)\) denotes the mean or optimal action according to a learned policy \(\pi\). \(\mathbf{P}(s)\) is a state-dependent and positive semi-definite square matrix that scales the stability of the advantage function and ensures efficient maximization of the Q-value function.

As noted, the advantage function \(A(s,a)\) is a concave quadratic function that decreases dramatically when an action \(a\) deviates from the optimal action learned \(\mu(s)\) and consequently maximizes the Q-value function \(Q_{\pi}(s,a)\). This characteristic provides an efficient way to evaluate the optimal action in multi-parametric continuous spaces, which is a key advantage of the NAF algorithm. Moreover, it ensures the exploration of new actions by injecting noise into the mean action (i.e., \(a = \mu(s) + n' \)) similarly to the exploration approach in the DDPG algorithm. This is typically done by using a Gaussian distribution around \( \mu(s)\), allowing the discovery of potentially better actions compared to the optimal action currently learned.

The key insight behind the NAF algorithm is the normalization of the advantage function, ensuring that it is consistently scaling across different states and actions. This helps stabilize learning and improves the robustness of the NAF algorithm compared to similar algorithms such as DDPG \cite{gu_continuous_2016}\cite{sun_distributed_2023}.

\begin{algorithm}
\caption{DRL-based NAF Learning Algorithm}\label{alg:cap}
\begin{algorithmic}[1]
\State $\text{Initialize Evaluation network } Q^E_{\pi}(s,a|\theta^Q_E)$
\State $\text{Initialize Target network } Q^T_{\pi}(s,a|\theta^Q_T)$
\Statex \hspace{\algorithmicindent} $\text{with weight } \theta^Q_T \gets \theta^Q_E $
\State $\text{Initialize replay buffer } D \gets \emptyset $
\For{$episode \gets 1, N_e$}
\State $\text{Initialize the environment and get the initial state } s_{t=0}$
 \State $\text{Initialize the OU random process }\mathcal{N} \; \text{to explore}$ 
\Statex \hspace{\algorithmicindent}\; $\text{ the actions}$
\For{$t \gets 1, N_t$}
\State $\text{Select a valid action } a_t = \mu(s_t|\theta^{\mu}) + \mathcal{N}_t$
\State $\text{Execute } a_t \text{, get a reward}\; r_t \;\text{and observe}$
\Statex \hspace{\algorithmicindent} $\text{ the next state} \;  s_{t+1}$
\State $\text{Store an experience } e_t = (s_t, a_t, r_t, s_{t+1})$
\Statex \hspace{\algorithmicindent} $\text{in the replay buffer } D$ 
\For {$n \gets 1, I$}
\State $\text{Sample a random mini-batch from } D$
\State $\text{Update the target network value } z_n \text{ as Eq.\ref{eq:target-value-update-fun}}$
\State $\text{Update the weight } \theta^Q_E ~ \text{by minimizing the loss } $
\Statex \hspace{\algorithmicindent} $\text{ function as in Eq.\ref{eq:loss-function}}$
\State $\text{Update target weight } \theta^Q_T \text{ as in Eq.\ref{eq:target-weight-update-fun}}$
\EndFor
\EndFor
\EndFor
\end{algorithmic}
\label{alg:DRL-NAF-training}
\end{algorithm}

\subsection*{DRL-based NAF Architecture}
Here, we discuss the implementation of the NAF algorithm to estimate the optimal Q-value function and the power allocation policy using the deep neural network architecture.  

{Conceptually, three artificial neural networks (ANNs) are required to implement the NAF algorithm, but they are implemented as a single shared network with three heads.} The first network approximates the advantage function \(A(s,a)\) that estimates how good or bad a particular action \(a\) is compared to the best action in a state \(s\). The second network maximizes the state-value function \(V(s)\) that estimates the expected cumulative reward starting from a state \(s\) under the optimal policy. The last network predicts the mean action \(\mu(s)\) (i.e., the policy function).

\begin{figure}[h!tbp]
    \centering
    \includegraphics[width=0.8\linewidth]{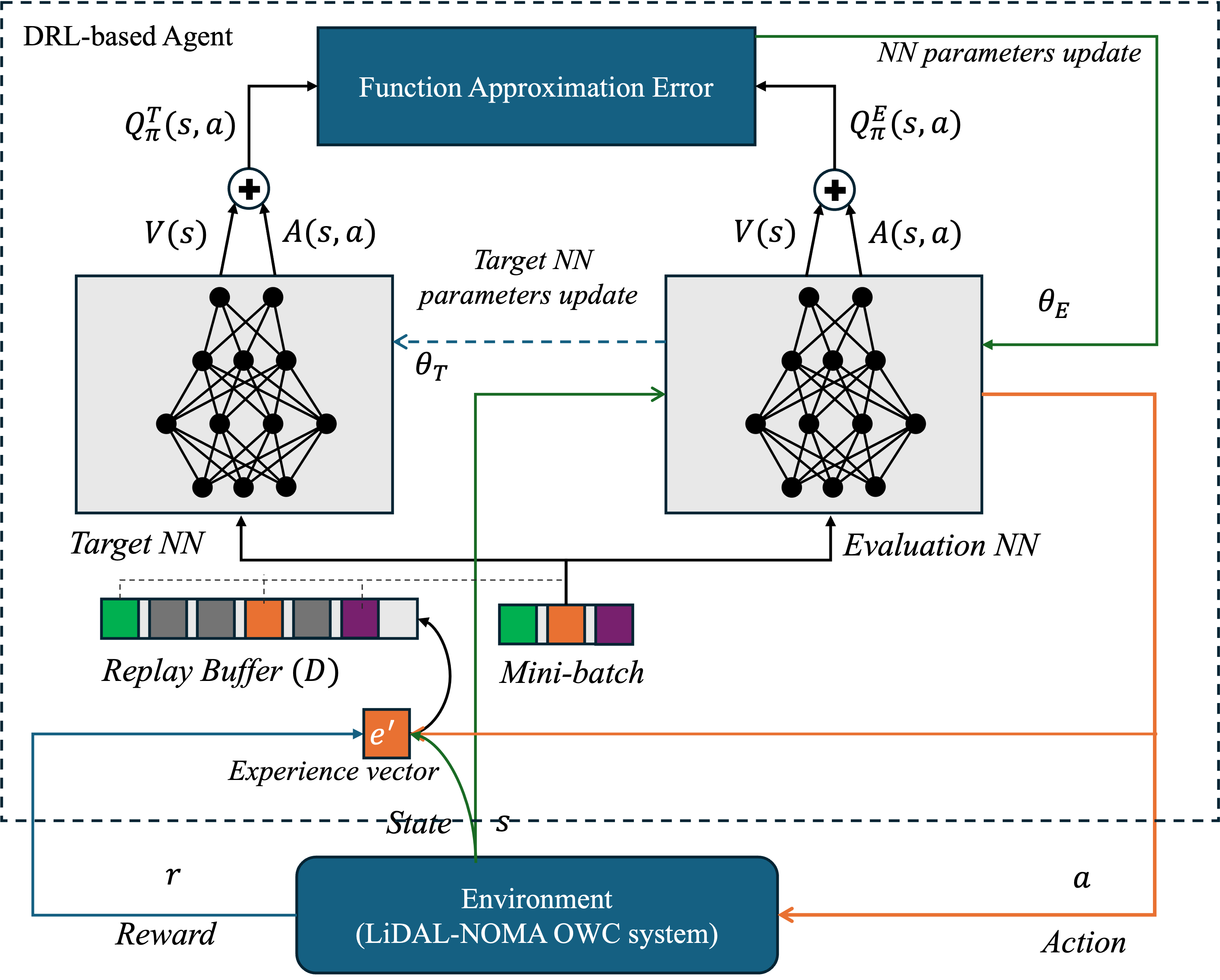}
    \caption{DRL-based NAF neural networks architecture.}
    \label{fig:DRL-NAF-architecture}
\end{figure}

Practically, these three ANNs can be combined into a single neural network architecture with shared hidden layers, as illustrated in Fig. \ref{fig:DRL-NAF-architecture}. This combined ANNs architecture generates three unique head outputs: a state vector that represents the current values of state \(s\), a continuous action vector \(a\) the RL agent should take in a given state \(s\), and a parameterised scaling matrix \(\textbf{P}(s)\). This integrated ANNs architecture allows efficient learning and representation of all components simultaneously, reducing computational overhead and learning time compared to DDPG architecture and conventional exhaustive search.

For the action exploration strategy, the Ornstein-Uhlenbeck process (OU) \cite{gu_continuous_2016} is proposed as a stochastic process that adds noise to a selected new action to revert or return to a long-term mean value \(\mu(s)\). This process helps to gradually reduce the noise of the exploration over time. As a result, the agent can refine and stabilize its actions as it learns. 
However, using a single DRL-based architecture with NAF can lead to the bootstrapping problem where the RL network uses its own predictions to update itself, which inevitably destabilizes the proposed RL algorithm and the reliability of the learning process. Therefore, an architecture of two identical neural networks is proposed and these two networks can be defined as follows: 
\begin{itemize}
    \item \textbf{Evaluation Network} that actively learns and approximates the Q-value and advantage functions and the policy based on the current interactions (i.e., experience) of the agent with the environment, as explained in the previous subsection. 
    \item \textbf{Target Network} provides a more stable estimate of the Q-value and advantage functions during learning updates (i.e., episodes) by comparing the difference between its Q-value and the one computed in a next state by the evaluation network (i.e., loss function), which is minimized by the decent gradient to update the weights of the evaluation network. The Q-value function of the target network can be described by \cite{gu_continuous_2016}
    \begin{align}
        z_n = r_n + \varsigma V_T(s_{n+1}| \theta^Q_T)
        \label{eq:target-value-update-fun}
    \end{align}
    where \(z_n\) is the Q-value function of the target network, \(n \in I\) denotes an iteration index of a small batch of random experiences in the replay buffer \(D\), and \(\theta_T\) denotes the weights of the target network.  

    Subsequently, the evaluation network weights are updated by applying the following loss function \cite{gu_continuous_2016}
    \begin{align}
        L = \dfrac{1}{I} \sum^I_{n=1} (z_n - Q^E_{\pi}(s_n,a_n|\theta^Q_E))^2
        \label{eq:loss-function}
    \end{align}
    Hence, \(I\) denotes the number of samples in the selected batch of experiences (i.e., mini-batch) and \(\theta^Q_E\) denotes the current weights of the evaluation network. 
    Then the target network weights are periodically updated using a soft update rule as expressed by \cite{gu_continuous_2016}
    \begin{align}
        \theta^Q_{T} \leftarrow \tau_Q \theta^Q_E + (1- \tau_Q) \theta^Q_T
        \label{eq:target-weight-update-fun}
    \end{align}
    where \(\tau_Q\) is a small factor.
\end{itemize}
As noted, for the learning process, a replay buffer \(D=s\{e_1,e_2,\cdots,e'\}\) is required to store the experiences that are sampled to train the current evaluation network and trigger updates to the target network during the learning intervals. In summary, the proposed DRL-based NAF architecture as illustrated in Fig.\ref{fig:DRL-NAF-architecture} stabilizes the learning process and efficiently converges to the optimal Q-value function.

% or
% do not use \section* anymore after \appendix, only \section**

% then use \section* to start each appendix
% you must declare a \section* before using any
% \subsection* or using \label (\appendices by itself

%
%\appendices

%\section*{}
\section*{Data Availability}

All relevant data and figures supporting the findings of the document are available on reasonable request. Please refer to Ahmed A. Hassan at ahmed.4.hassan@kcl.ac.uk. 

\section*{Acknowledgments}

The authors acknowledge funding from the Engineering and Physical Sciences Research Council (EPSRC), in part by the TOWS project under grant (EP/S016570/2) and in part by the TITAN project under grant (EP/X04047X/2). For the purpose of open access, the authors have applied a Creative Commons Attribution (CC BY) license to any author accepted manuscript version arising.

 \section*{Author Contributions}
AH developed the LiDAL-Assisted RLNC-NOMA system model and implemented it in MATLAB. He also formulated the power allocation problem and designed and implemented machine learning algorithms using Python in MATLAB. AH also conducted all simulations, performance evaluations, and generated results. AQ contributed to the system and mathematical modeling, reviewed the results, and assisted in the preparation of the manuscript. TE reviewed the model and the results and supported in the preparation of the manuscript. JE is the originator of the main research idea and contributed to the preparation and refinement of the manuscript. All authors reviewed and approved the final version. 

\section*{Competing Interests}

The authors declare no competing financial or non-financial interests.

\ifCLASSOPTIONcaptionsoff
  \newpage
\fi




\makeatletter
\def\bstctlcite{\@ifnextchar[{\@bstctlcite}{\@bstctlcite[@auxout]}}
\def\@bstctlcite[#1]#2{%
  \@bsphack
  \@for\@citeb:=#2\do{%
    \edef\@citeb{\expandafter\@firstofone\@citeb}%
    \if@filesw
      \immediate\write\csname #1\endcsname{\string\citation{\@citeb}}%
    \fi
  }%
  \@esphack
}
\makeatother

\bibliographystyle{IEEEtran}
%\bibliographystyle{unsrt}
\bibliography{References/references}

\section*{Figure Legends}
\addcontentsline{toc}{section}{Figure Legends}

\noindent \textbf{Fig. 1.} \textbf{ Diagram illustrates an indoor study environment equipped with a LiDAL-based RLNC-NOMA system.} In this environment model, there are eight homogenized RGB-LD transmitters, each integrated with a LiDAL-based receiver, which collectively provide functionality for illumination, user localization, and RLNC-NOMA-based optical wireless communication.

\noindent \textbf{Fig. 2.} \textbf{Comparison between NAF and DDPG agents with respect to convergence of the learning process toward the optimal average reward \(\gamma\)}. The red  and blue curves indicate the convergence of both target and evaluation  neural networks (NNs) of the NAF agent. The yellow curve represents the convergence of DDPG actor-critic agent over training iterations. 

\noindent \textbf{Fig. 3.} \textbf{Comparison of the loss function associated with the training iterations of NAF and DDPG.} The graph shows the mean squared error between the predicted Q-values and the target Q-values for the NAF evaluation (red curve) and target (blue curve) networks compared to the DDPG (yellow curve).

\noindent \textbf{Fig. 4.} \textbf{Comparison of the achievable average sum rate across the eight LiDAL-based RLNC-NOMA user groups.} The graph illustrates the sum data rate achieved in gigabits per second (Gbps) by the proposed NAF (blue bars) in relation to the exhaustive search  benchmark (red bars), DDPG (yellow bars) and the conventional GRPA method (purple bars).

\noindent \textbf{Fig. 5.} \textbf{Comparison of the average user data rate for each LiDAL-based RLNC-NOMA group.} The figure highlights the average data rate per user in megabits per second (Mbps) in groups of four users connected to a single optical AP, where the power allocation factors selected by the proposed NAF scheme (blue circles), exhaustive search (red circles), DDPG (yellow circles), and GRPA (purple circles).

\noindent \textbf{Fig. 6.} \textbf{Jain's fairness index of user data rates per LiDAL-based RLNC-NOMA group.} The degree of user fairness, quantified by Jain’s fairness index, that achieved by the proposed NAF scheme (blue circles), the exhaustive search benchmark (red circles), the DDPG-based approach (yellow circles) and the GRPA scheme (purple circles).

\noindent \textbf{Fig. 7.} \textbf{Average power allocation as a function of user location estimation error in the LiDAL-based RLNC-NOMA group.} The figure illustrates how the average power allocation factor \(\alpha\) varies with the estimated location error in meters (m) for the proposed NAF scheme (blue bars), the exhaustive search (red bars), the DDPG-based method (yellow bars), and the GRPA scheme (purple bars).

\noindent \textbf{Fig. 8.} \textbf{Schematic diagram of the RL framework.} The diagram illustrates how an agent interacts with an environment over time: at each step  \(t\), the agent observes the current state \(s_t\), selects an action \(a_t\) according to its policy, and receives feedback from the environment in the form of a reward \(r_t\) and a new state. 

\noindent \textbf{Fig. 9.} \textbf{Illustrative diagram of the architecture of the DRL-based NAF agent neural networks.} The diagram presents the components of two identical neural networks (NNs) that utilize a shared replay buffer \(D\) batch and synchronized weight updates \(\theta\) while interacting with the LiDAL-based RLNC-NOMA system model to optimize power allocation according to the objective of the defined optimization problem. The  approximation error function is employed to evaluate the Q-values predicted by both NNs to optimize their weights. 

\section*{Table Legends}
\addcontentsline{toc}{section}{Table Legends}

\noindent \textbf{TABLE I.} \textbf{DRL-based NAF agent hyperparameters.}

\noindent \textbf{TABLE II.} \textbf{Implementation performance benchmark.}

\section*{Algorithm Legends}
\addcontentsline{toc}{section}{Algorithm Legends}

\noindent \textbf{Algorithm 1.} \textbf{DRL-based NAF learning algorithm.}
\end{document}